\documentclass[aps,prl,twocolumn,superscriptaddress]{revtex4}

\usepackage{graphicx}
\usepackage{dcolumn}
\usepackage{bm}
\usepackage{amsmath}
\usepackage{amssymb}
\usepackage{latexsym}
\usepackage{epsfig}
\usepackage{amsbsy}
\usepackage{array}
\usepackage{amssymb}
\usepackage{setspace}
\usepackage{bm}
\usepackage{color}

\def\sint{\ifmmode{- \!\!\!\!\!\! \int}
    \else{\hbox{$- \!\!\!\! \int \ $}}\fi}

\begin{document}

\newskip\beforecaptionskip
\newskip\aftercaptionskip
\beforecaptionskip=0pt 
\aftercaptionskip=0pt

\preprint{Physical Review Letters}

\title{All-optical control of the spontaneous emission of quantum dots \\ using coupled-cavity quantum electrodynamics}

\author{C.~Y.~Jin \footnote{Electronic mail: c.jin@tue.nl}}
\author{M.~Y.~Swinkels \footnote{These two authors contributed equally to this work}}
\author{R.~Johne $^{\dagger}$}
\author{T.~B.~Hoang}
\author{L.~Midolo}
\author{P.~J.~van~Veldhoven}
\author{A.~Fiore}
\affiliation{COBRA Research Institute, Eindhoven University of Technology, P.O.Box 513, 5600MB Eindhoven, the Netherlands}

\date{\today}

\begin{abstract}
We demonstrate the remote all-optical control of the spontaneous emission (SE) of quantum dots using coupled photonic crystal cavities. By spectrally tuning a Fabry-Perot cavity in resonance with a target cavity, the quality factor and the local density of states experienced by emitters in the target cavity are modified, leading to a change in the SE rate. From the theoretical analysis of the coupled-cavity quantum electrodynamics system, the SE rate change can be higher than the quality factor change due to a reduction of the vacuum field at the emitter\textquoteright s position when the two cavities are brought in resonance. Both the weak and strong coupling regimes of two cavities have been observed experimentally and the SE decay rate has been modified by more than a factor of three with remote optical control.  
\end{abstract}



\maketitle


Semiconductor quantum dots (QDs) precisely positioned inside photonic crystal (PhC) cavities represent a scalable platform for solid-state cavity quantum electrodynamics (CQED) \cite{Vahala2003, Yoshie2004, Kimble2008, OBrien2009}. In general, quantum information should be stored in the charge or spin degree of freedom in the quantum system, and then released to a photonic channel, e.g. through spontaneous emission (SE). The SE rate can be greatly enhanced by increasing the local density of states (LDOS) at the emitter\textquoteright s position using a wavelength-sized, low-loss cavity \cite{Purcell1946,Yablonovitch1987, Sprik1996}. In order to control the timing of the emission process, a procedure is needed to switch the emitter-cavity coupling on and off. An attractive approach will be the ultrafast control of the cavity characteristics, leading to a dynamic modulation of the LDOS and therefore of the SE rate. The dynamic tuning of the quality factor (Q-factor) recently demonstrated in silicon-based PhC cavities is an example of such cavity modulation techniques \cite{Tanaka2007, Tanabe2009}. A static and quasi-permanent change of the Q-factor and of the emission intensity of QDs was recently observed by producing a structural modification in the vicinity of the cavity \cite{Nakamura2011}. 

However, for the practical application in CQED, it requires a remote and reversible control scheme which produces large changes in the LDOS but does not affect the emitter-cavity interaction by local perturbations. The coupled-cavity quantum electrodynamics (CCQED) platform proposed by Hughes \cite{Hughes2007} is an interesting candidate for such remote and reversible control, and several demonstrations of strong coupling between PhC cavities have been recently reported \cite{Atlasov2008, Vignolini2009, Sato2011, Dundar2012}. In this work, we present the experimental observation of the controlled and reversible modification of the SE rate in a {\textquotedblleft target\textquotedblright} cavity by tuning the properties of a remote {\textquotedblleft control\textquotedblright} cavity. The tuning is implemented using thermo-optic effects, but the concept is readily extendable to the ps time scale by using free-carrier injection, opening the way to the dynamic control of SE in solid-state cavities.

The enhancement of the SE rate $\gamma$ in a cavity as compared to the SE rate in the bulk $\gamma_0$ is generally described in perturbation theory by \cite{Purcell1946, Andreani1999},
\vspace{0pt}
\begin{eqnarray}
\setlength{\abovedisplayskip}{1pt}
\setlength{\belowdisplayskip}{1pt}
\dfrac{\gamma}{\gamma_{0}}&=&\left(\dfrac{3\lambda^3}{4\pi^2n_r^3}Q\left|\mathbf{E(r_0)}\right|^2\right)\cdot\dfrac{\left(\delta\lambda/2\right)^{2}}{\left(\delta\lambda/2\right)^{2}+\left(\lambda-\lambda_{0}\right)^2}
\label{eq:purcell}
\end{eqnarray}	
					
\noindent where $\lambda$ is the cavity mode wavelength, $\lambda_0$ is the dipole emission wavelength,  $\delta \lambda$ is the linewidth of the cavity mode, $n_r$ is the refractive index, $Q$ is the Q-factor of the cavity mode and $\mathbf{E(r_0)}$ is the normalized mode function at the dipole position $\mathbf{r_0}$. The first term at the right-hand side of Eq. (\ref{eq:purcell}) is the well-known Purcell factor, $F$, which is governed by the Q-factor and the mode function distribution or inversely the mode volume. The second term describles the effect of the wavelength detuning. The SE rate is conventionally controlled through the wavelength detuning, e.g. by changing the emitter's energy or the cavity mode frequency. Instead, in the following we demonstrate a remote optical control of the SE rate by a real-time change of the Purcell factor $F$ using coupled cavities.

\begin{figure}[!htb]
\centering
\vspace{0pt}
\begin{minipage}{0.5\textwidth}
\centering
\includegraphics[width=0.8\linewidth]{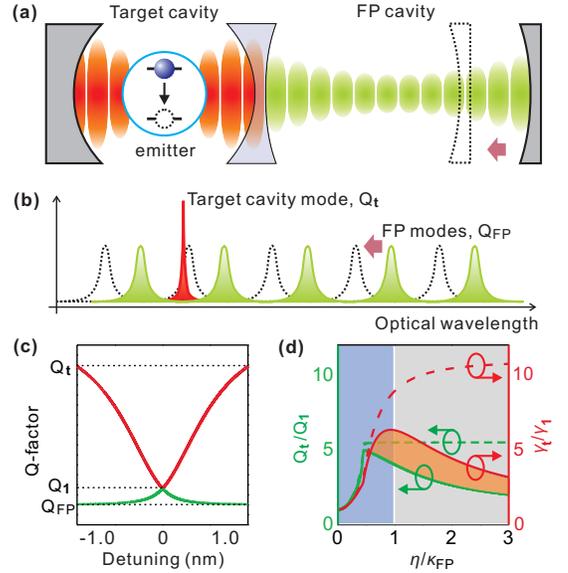}
\end{minipage}
\caption{(a) A schematic picture of a CCQED system. An emitter is placed in the target cavity with a higher Q-factor. A control cavity (FP cavity) which has periodic modes with lower Q-factors is optically coupled to the target cavity through a semitransparent mirror. (b) By controlling the effective cavity length of the FP cavity, FP modes can be tuned in resonance with the target cavity mode. (c) Calculated Q-factors for the coupled modes in the strong coupling regime as a function of the wavelength detuning between the original cavity modes. (d) Calculated Q-factor change (green curves) and the SE rate change (red curves) as a function of the normalized coupling strength. Dashed curves are for the ideal case of a single FP mode and solid curves take into account multiple FP modes. The blue and grey area indicates the weak and strong coupling regime between two cavities, respectively. The orange shadow highlights where the SE rate change can be higher than the Q-factor change due to the field delocalization. \label{fig:Graph1}}
\end{figure}

Fig. \ref{fig:Graph1} (a) shows an ideal emitter-cavity scheme where one of the cavity mirrors is replaced by a Fabry-Perot (FP) etalon. The FP cavity mode has a periodic mode structure sketched in Fig. \ref{fig:Graph1} (b) and it is assumed to have a lower Q-factor compared to the target cavity mode. By adjusting the length of the FP cavity or modulating the refractive index of the media inside it, the set of FP modes is spectrally tuned with respect to the target cavity mode. At resonance the coupling causes an additional leakage for the optical field confined in the target cavity resulting in a change of the Q-factor and the mode function distribution $\mathbf{E(r)}$. Hence, the SE rate of emitters in the target cavity can be controlled non-locally by adjusting the resonant condition. Differently from the structure proposed in Ref. \cite{Hughes2007}, we use evanescent coupling between two adjacent cavities and an extended FP cavity for the control, whose periodic mode structure facilitates the spectral resonance with the target cavity.   

The presented system is described by a three-oscillator model including the dipole, the target cavity and the FP cavity. The dressed states of the system are obtained by diagonalizing the non-Hermitian Hamiltonian \cite{CohenTannoudji1998} 
\vspace{0pt}
\begin{eqnarray}
\bf{H}&=&\hbar
\left[
\begin{array}{ccc}
    \omega_0 & g & 0 \\
    g &  \omega_t-i\kappa_t & \eta \\
    0 &  \eta & \omega_{FP}-i\kappa_{FP} \\
\end{array}
\right]
\label{eq:hamiltonian}
\end{eqnarray}		

\noindent where $\omega_0$, $\omega_t$ and $\omega_{FP}$ denote the angular frequencies of the dipole, the target cavity mode and the FP mode, the corresponding losses are related to the Q-factor, $\kappa=\omega/2Q$, and $\eta$ is the coupling rate between the two cavities.  The dipole is placed in the target cavity and spatially decoupled from the FP modes. The weak coupling between the dipole and the target cavity ($g \ll \kappa_t, \kappa_{FP}$) is always assumed and the terms of {\textquotedblleft strong coupling\textquotedblright} and {\textquotedblleft weak coupling\textquotedblright} in the following discussion are only related to the coupling between cavities. We first solve the subsystem of coupled cavities (see Supplementary Materials), which yields the complex eigenvalues
\vspace{0pt}
\begin{eqnarray}
\begin{aligned}
\tilde{\omega}_{1,2}=&\dfrac{\left(\omega_t-i\kappa_t\right)+\left(\omega_{FP}-i\kappa_{FP}\right)}{2} \\
&\pm\dfrac{1}{2}\sqrt{\left(\left(\omega_t-i\kappa_t\right)-\left(\omega_{FP}-i\kappa_{FP}\right)\right)^2+4\eta^2}.
\end{aligned}
\label{eq:eigen}
\end{eqnarray}

\noindent The effect of wavelength detuning $\lambda_t-\lambda_{FP}$ on the Q factor of the coupled cavity modes is illustrated in Fig. \ref{fig:Graph1} (c). By tuning a low Q cavity mode through a high Q target cavity mode a significant change in the imaginary parts of the coupled eigenfrequencies $\mathtt{Im}(\tilde{\omega}_{1,2})$ hence in the Q-factor can be observed at the crossing point.

Transforming the matrix (\ref{eq:hamiltonian}) into the basis of the coupled cavity modes in Eq. (\ref{eq:eigen}) yields  
\vspace{0pt}
\begin{eqnarray}
\begin{split}
\bf{H'} = \hbar\left[
\begin{array}{ccc}
    \omega_0 & \alpha g & \beta g \\
    {\alpha g} &  \tilde{\omega}_1 & 0 \\
    {\beta g} &  0 & \tilde{\omega}_2 \\
\end{array}
\right].
\label{eq:Hbase}
\end{split}
\end{eqnarray}		

\noindent The normalized mode functions, $\mathbf{E_{1,2} (r)}$, of the coupled-cavity system are mixtures of the eigenfunctions of the initial cavity modes, $\mathbf{E}_{1,2}\mathbf{(r)}=\alpha_{1,2}\mathbf{E}_t\mathbf{(r)}+\beta_{1,2}\mathbf{E}_{FP}\mathbf{(r)}$, where $\alpha_1=\beta_2=\alpha$, $\alpha_2=-\beta_1=\beta$, and $\left|\alpha\right|^2+\left|\beta\right|^2=1$. Consequently, the coupling constants between the dipole and the new eigenmodes are weighted by their target cavity fraction due to the positioning of the dipole. Because of the mode coupling, the mode distribution of the state $\tilde{\omega}_{1}$ becomes increasingly delocalized for decreasing detuning, leading to a reduction of the vacuum field at the dipole\textquoteright s position. This effect can be intuitively interpreted as a change of the effective mode volume due to the field delocalization.

The modulation of the Q-factors and the mode function distribution will affect the dynamics of the dipole-cavity interaction. For instance in the case that the two coupled cavities are in the strong coupling regime $(\eta \gg \kappa_c,\kappa_{FP})$, due to the energy splitting a narrow dipole will interact with only one of the two eigenmodes,  leading to $\gamma_1 \cong \left| \alpha \right|^2 \dfrac{Q_1}{Q_t} \gamma_t= \dfrac{1}{2} \dfrac{Q_1}{Q_t} \gamma_t$, where $Q_t$ is the uncoupled Q-factor of the target cavity, $Q_1$ is the higher Q-factor of coupled modes at resonance, and $\gamma_t$ and $\gamma_1$ are the corresponding SE rates. It follows that the SE ratio between the off-resonant and on-resonant cases, $\gamma_t/\gamma_1$, is not only affected by the Q-factor ratio, $Q_t/Q_1$, but also enhanced by a factor of two due to the mode delocalization. In contrast, in the weak coupling $(\eta \ll \kappa_c,\kappa_{FP})$, the SE enhancement has contributions from both coupled modes $\gamma \cong \left| \alpha \right|^2 \dfrac{Q_1}{Q_t} \gamma_t+\left| \beta \right|^2 \dfrac{Q_2}{Q_t} \gamma_t$, which yields $\gamma \cong \gamma_t$ in the case of $Q_1=Q_t$.

\begin{figure}[!htb]
\vspace{0pt}
\centering
\begin{minipage}[b]{0.5\textwidth}
\centering
\includegraphics[width=0.85\linewidth]{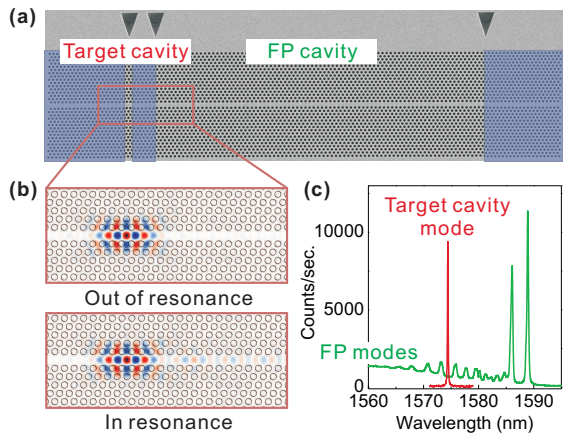}
\end{minipage}
\caption{(a) A SEM image of the fabricated PhC cavities. Three blue-shadowed squares indicate the barriers of waveguide cavities. (b) Simulated electrical field (y-polarized, $E_y$) distribution between two cavities when they are out of resonance (top) and in resonance (bottom). (c) $\mu$PL spectra of CCQED system when focusing the excitation beam on the target cavity (red) and FP cavity (green). \label{fig:Graph2}}
\end{figure}

Fig. \ref{fig:Graph1} (d) presents simulation results for the CCQED system using matrix (\ref{eq:hamiltonian}). We set $Q_{t} =2 \times 10^4$, $Q_{FP}=2 \times 10^3$, and the dipole is always resonant with the coupled mode which has higher Q-factor \cite{Note}. The Q-factor ratio, $Q_{t}/Q_{1}$, is plotted as a function of the coupling strength divided by the FP cavity loss, $\eta / \kappa_{FP}$ (red dashed curve). The corresponding ratio of the SE decay rate, ${\gamma_{t}}/{\gamma_{1}}$, of the dipole as a function of the coupling strength is also plotted (green dashed curve). While in the weak coupling regime the SE change follows the change in the Q-factor, in the strong coupling regime the SE change is up to a factor of two larger than the Q-change as discussed above. In reality, the target cavity mode can couple to more than one FP mode simultaneously in case of high coupling strength. The Q-factor change and the SE change of the dipole with multiple FP modes are shown by the green and red solid curves in Fig. \ref{fig:Graph1}(d). Three equally spaced FP modes with 3 nm separation are taken into account in simulation. Due to the more complex interaction of multiple modes the Q-factor change and the change in the spontaneous emission rate are modified with respect to the single FP mode calculation. However, the above analysis of the decay rates for the weak and strong coupling still holds.

\begin{figure}[!htb]
\vspace{0pt}
\centering
\begin{minipage}{0.5\textwidth}
\centering
\includegraphics[width=1\linewidth]{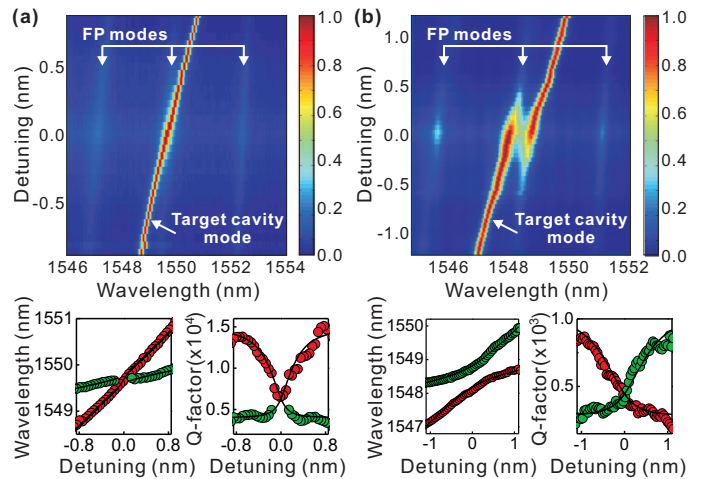}
\end{minipage}
\caption{(a) Weak and (b) strong coupling between two PhC cavities in a single-beam measurement on the target cavity. The spectral maps are constructed from spectra with normalized intensity.} The continuous black lines are fits based on the coupled mode theory. \label{fig:Graph3}
\end{figure}

In the experiment, our platform for the CCQED is based on PhC cavities using QDs as the emitter. Cavities were realized within a hexagonal PhC lattice fabricated in an InGaAsP membrane which contains a single layer of self-assembled InAs QDs with a dot density around $2 \times 10^9 cm^{-2}$ (see Supplementary materials). The QD ensemble gives a broad emission centered at 1570nm \cite{Anantathanasarn2005}. Fig. \ref{fig:Graph2} (a) shows the scanning electron microscope (SEM) image of the fabricated structure based on two local defects along a W1 PhC waveguide. The target cavity is constituted by a high-Q double-heterostructure (DHS) cavity \cite{Song2005} with two periods of the lattice constant slightly larger than the original lattice constant, $a_1=1.03 \times a_0$. The FP cavity consists of eighty periods of modulated lattice constant ($a_2=a_1$). Fig. \ref{fig:Graph2} (b) presents the simulated electric field distribution based on 3D FDTD method \cite{MEEP}. A leakage from the target cavity to the FP cavity can be clearly observed when the two cavities are in resonance. The measurements were performed in a confocal microscope at a temperature of 77 K, in order to reduce the influence from the non-radiative recombination and the homogenous broadening of QDs, while allowing the thermo-optic tuning of the PhC cavity \cite{Fushman2007}. A diode laser emitting at 780 nm was used for exciting the cavity and the $\mu$PL signal collected by the objective is measured with a spectrometer and an InGaAs array. Fig. \ref{fig:Graph2}(c) shows two typical $\mu$PL spectra taken from single-beam measurements by focusing the laser beam on either the target or the FP cavity. The FP cavity has a quasi-periodic mode structure, which is governed by the modified dispersion relation close to the slow-light frequency of the PhC waveguide \cite{Hoang2012}.

Firstly, single-beam measurements have been performed with one laser beam focused on the target cavity and the signal collected from the same position. By increasing the laser power, a red shift of the target cavity mode due to thermo-optic tuning is observed with only a small shift of FP modes caused by heat diffusion \cite{Dundar2012}. Fig. \ref{fig:Graph3}(a) shows a weak coupling case in a PhC with 4.5-period barrier between two cavities and 20-period barriers on the two sides. The main panel in Fig. \ref{fig:Graph3}(a) shows the spectral map with normalized intensities for various detunings, while the two small plots present the cavity mode wavelength and Q-factor as functions of the wavelength detuning. It exhibits a clear crossing of two coupled modes and indicates the weak coupling regime. The Q-factor of the target cavity decreases down from 13800 to 6500 and then increases again to 15100, whilst the Q-factor of the FP cavity increases slightly from 4000 to 4900 and then decreases to 3500. By fitting the crossing behavior of two coupled modes using Eq. \ref{eq:eigen}, as shown by the black lines, we extract the parameters $Q_t=1.5 \times 10^4$, $Q_{FP}=4 \times 10^3$, and $\eta=0.8 \times \kappa_{FP}$. Fig. \ref{fig:Graph3}(b) exhibits a different situation where the barrier thickness between two cavities is reduced to 4 periods. In this case, two coupled modes shows anti-crossing, which indicates the strong coupling regime. The Q-factor of the mode at shorter (longer) wavelength varies from 8500 to 1800 (from 2600 to 8500), respectively. Numerical fitting reveals $Q_t=1.1 \times 10^4$, $Q_{FP}=2.5 \times 10^3$, and $\eta=1.1 \times \kappa_{FP}$. 

\begin{figure}[!htb]
\vspace{0pt}
\centering
\begin{minipage}{0.5\textwidth}
\centering
\includegraphics[width=0.95\linewidth]{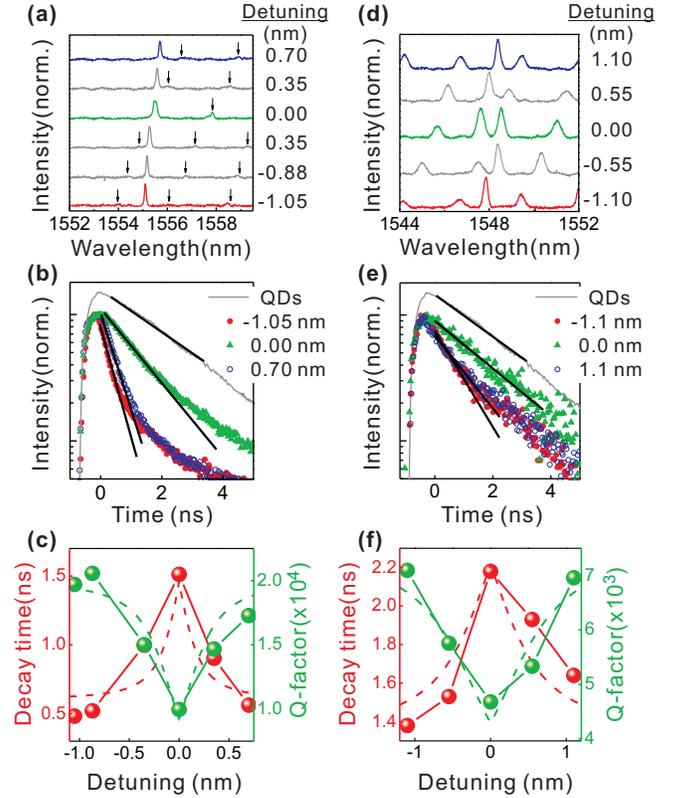}
\end{minipage}
\caption{Remote control of the Q factor and SE rate using two beam measurement for a weak coupling case (a-c) and a strong coupling case (d-f). (a) and (d) $\mu$PL spectra measured at different wavelength detuning. (b) and (e) SE decay curves at different wavelength detuning. The grey curve is the SE decay from QDs in the bulk.  (c) and (f) The decay time and Q-factor at the high-Q mode as a function of the wavelength detuning. The dashed curves are simulations based on the coupled mode theory. \label{fig:Graph4}}
\end{figure}

Two-beam measurements have been carried out to demonstrate the remote control of both Q-factor and SE rate with one pulsed laser beam ($\lambda=750 nm$, pulse width ~100 ps) focused on the target cavity and a CW heating laser beam focused at the center of the FP cavity, which is tilted slightly of the optical axis of the objective lens in order to position it at a distance of a few tens of micrometers from the other beam. Fig. \ref{fig:Graph4}(a) presents the spectra with weakly-coupled cavities at different pump powers of the tuning laser. The decay time of QDs resonant with the high-Q coupled mode has been measured by time-correlated single photon counting with a superconducting single photon detector \cite{Zinoni2007}, using a narrow bandpass filter ($\Delta \lambda=0.5nm$). Fig.  \ref{fig:Graph4}(b) shows the SE decay at the high-Q mode frequency with different wavelength detunings between the cavities and the decay of the QD emission in the bulk (outside the PhC), in a $\Delta \lambda=12 nm$ bandwidth, as a reference (grey curve). By fitting the initial part of the curve with an exponential decay function, the spontaneous emission lifetime modulated by PhCs is obtained, while the slow decay component is attributed to dark excitons \cite{Johansen2010}. Compared to the 2.5 ns decay time of the QD emission, the decay time at target cavity is 0.5 ns and 1.5 ns when two weakly coupled modes are off-resonance and on-resonance, respectively. The strong coupling case has been also studied as shown in Fig. \ref{fig:Graph4}(d)-(f). The SE decay time is 1.4 and 2.2 ns for the high-Q mode, when two cavity modes are off-resonance and on-resonance, respectively.

To fit the experimental decay times we use the expression $\gamma^{exp}=\gamma_{cav}^{theor}+\gamma_{leak}$, where $\gamma_{cav}^{theor}$ is the total emission rate into the two coupled modes calculated from diagonalizing the matrix (\ref{eq:hamiltonian}), while $\gamma_{leak}=(3.1ns)^{-1}$ is the emission rate into the out-of-plane leaky modes which is measured from the decay time of QDs in the PhC mirror region. The global fit as shown by the dashed curves in Fig. \ref{fig:Graph4}(c) and (f) closely matches the experimental results. $\eta/\kappa_{FP}=0.95$ and 1.40 can be identified for the weak and strong coupling, respectively. Both are close to the transition regime defined by  $\eta/\kappa_{FP}=1$. The measured decay rate into the cavity modes changes by a factor of 3.7 while the Q factor changes by a factor of 2.1 in the weak coupling regime. In the strong coupling case the cavity decay rate changes by a factor of 2.7 while the Q factor changes by a factor of 1.6. Both cases confirm that in case $\eta/\kappa_{FP}>0.5$ the SE rate change can be higher than the Q-factor change. 

In summary, we have demonstrated the real-time, remote all-optical control of the SE rate by more than a factor of three in a coupled PhC cavity system where the control beam is separated more than 20 $\mu$m from the target cavity. The results have been interpreted in the framework of coupled mode theory, showing that the SE rate is affected by the change in both the Q-factor and the mode volume. These results open the way to the ultrafast control of the QD-cavity interaction by replacing the thermo-optic tuning with free-carrier injection on the ps time scale.

\vspace{10pt}

\begin{acknowledgements}
The authors are grateful to B. Wang, M. A. D$\rm{\ddot{u}}$ndar, and R. W. van der Heijden for fruitful discussions, and to E. J. Geluk, E. Smalbrugge, and T. de Vries for technical support. This research is supported by the Dutch Technology Foundation STW, applied science division of NWO, the Technology Program of the Ministry of Economic Affairs under project No. 10380, the FOM project No. 09PR2675, and NanoNextNL, a micro and nanotechnology program of the Dutch ministry of economic affairs, agriculture and innovation (EL\&I) and 130 partners.
\end{acknowledgements}


\begin{thebibliography}{35}
\bibitem{Note}
It should be noted that, in the case of a dipole with fixed energy, the SE rate change in the strong coupling regime will be even higher due to the spectral decoupling between the dipole and coupled mode as the latter is tuned. However, in the experiments described in this work the SE rate of emitters resonant with the coupled mode is measured and the results in Fig. 1 apply.
\end{thebibliography}


\section{Supplementary materials}
\setcounter{equation}{0}
\renewcommand{\theequation}{S.\arabic{equation}}

\subsection{\leftline{1.Oscillator model for the coupled cavity system}}

The three-oscillator model for the dipole, the target cavity and the FP cavity can be described by a non-Hermitian Hamiltonian
\vspace{0pt}
\begin{eqnarray}
\bf{H}&=&\hbar
\left[
\begin{array}{ccc}
    \omega_0 & g & 0 \\
    g &  \omega_t-i\kappa_t & \eta \\
    0 &  \eta & \omega_{FP}-i\kappa_{FP} \\
\end{array}
\right],
\label{eq:hamiltonian_S}
\end{eqnarray}		
where $\omega_0$, $\omega_t$ and $\omega_{FP}$  are the angular frequencies of the dipole, the target cavity mode and the FP mode, respectively. $\kappa_t$ and $\kappa_{FP}$ describe the intrinsic linewidths of the uncoupled cavity modes, which are related to the Q-factor, $\kappa=\omega/2Q$. The dipole is assumed to be spatially decoupled from the FP modes and is coupled to the target cavity with a coupling rate of $g$. The coupling rate between the target cavity and the FP cavity is described by $\eta$. 

At first we consider the submatrix which contains the components from the target and FP cavity
\vspace{0pt}
\begin{eqnarray}
\it{\bf{H_s}}&=&\hbar
\left[
\begin{array}{ccc}
    \omega_t-i\kappa_t & \eta \\
    \eta & \omega_{FP}-i\kappa_{FP} \\
\end{array}
\right].
\label{eq:submatrix_S}
\end{eqnarray}	
By solving the eigenvalue equation $\Vert {\bf{H_s}}-\hbar \omega {\bf{I}} \Vert = 0$, the complex eigenvalues $\tilde{\omega}_{1,2}$ are obtained
\vspace{0pt}
\begin{eqnarray}
\begin{aligned}
\tilde{\omega}_{1,2}=&\dfrac{\left(\omega_t-i\kappa_t\right)+\left(\omega_{FP}-i\kappa_{FP}\right)}{2} \\
&\pm\dfrac{1}{2}\sqrt{\left(\left(\omega_t-i\kappa_t\right)-\left(\omega_{FP}-i\kappa_{FP}\right)\right)^2+4\eta^2}.
\end{aligned}
\label{eq:eigen_S}
\end{eqnarray}
The eigenvectors are found as 
\vspace{0pt}
\begin{eqnarray}
\left[
\begin{array}{ccc}
\alpha_1 \\
\beta_1 \\
\end{array}
\right]=&
\left[
\begin{array}{ccc}
    \dfrac{\eta}{\sqrt{\eta^2+\left|\tilde{\omega}_1-\left(\omega_t-i\kappa_t\right)\right|^2}} \\
    \dfrac{\tilde{\omega}_1-\left(\omega_t-i\kappa_t\right)}{\sqrt{\eta^2+\left|\tilde{\omega}_1-\left(\omega_t-i\kappa_t\right)\right|^2}} \\
\end{array}
\right], \\
\left[
\begin{array}{ccc}
\alpha_2 \\
\beta_2 \\
\end{array}
\right]=&
\left[
\begin{array}{ccc}
    -\dfrac{\tilde{\omega}_1-\left(\omega_t-i\kappa_t\right)}{\sqrt{\eta^2+\left|\tilde{\omega}_1-\left(\omega_t-i\kappa_t\right)\right|^2}} \\
    \dfrac{\eta}{\sqrt{\eta^2+\left|\tilde{\omega}_1-\left(\omega_t-i\kappa_t\right)\right|^2}} \\
\end{array}
\right], 
\label{eq:eigenfunction_S}
\end{eqnarray}	
where $\alpha_1=\beta_2=\alpha$, $\alpha_2=-\beta_1=\beta$, and $\left|\alpha\right|^2+\left|\beta\right|^2=1$. 

Going back to the three-oscillator case, the vectors used to diagonalize the submatrix read
\vspace{0pt}
\begin{eqnarray}
\begin{split}
A_1=
\left[
\begin{array}{ccc}
1 \\
0 \\
0 \\
\end{array}
\right], 
A_2= \left[
\begin{array}{ccc}
0 \\
\alpha \\
-\beta\\
\end{array}
\right], 
A_3 =  \left[
\begin{array}{ccc}
    0 \\
    \beta \\
    \alpha \\
\end{array}
\right]. 
\label{eq:basis_S}
\end{split}
\end{eqnarray}	
The transformation matrix can be constructed with the vectors in Eq. (\ref{eq:basis_S}), which changes the Hamitonian to the new bases of coupled cavity modes  
\vspace{0pt}
\begin{eqnarray}
\bf{T}&=&\left[
\begin{array}{ccc}
   1 & 0 & 0 \\
    0 & \alpha & \beta \\
    0 & -\beta & \alpha \\
\end{array}
\right].
\label{eq:Tmatrix_S}
\end{eqnarray}		
Thus, the Hamiltonian with the new bases as also shown in Eq. (4) can be derived
\vspace{0pt}
\begin{eqnarray}
\begin{split}
\bf{H'} = \bf{T^{-1}}\bf{H}\bf{T} = \hbar \left[
\begin{array}{ccc}
    \omega_0 & \alpha g & \beta g \\
    {\alpha g} &  \tilde{\omega}_1 & 0 \\
    {\beta g} &  0 & \tilde{\omega}_2 \\
\end{array}
\right].
\label{eq:hamiltonian2_S}
\end{split}
\end{eqnarray}

We note that the effect of pure dephasing and thus off-resonant cavity feeding are not taken into account in the above simple model. In principle, for large dephasing rates $\gamma^*$, a similar analysis as in the main paper can be carried out using a generalized Purcell formula for a broad emitter, which has been derived elsewhere \cite{Auffeves2010}. This leads to a similar expression in case of strong coupling for
\vspace{0pt}
\begin{eqnarray}
\setlength{\abovedisplayskip}{1pt}
\setlength{\belowdisplayskip}{1pt}
\gamma_1 \cong \left| \alpha \right|^2 \dfrac{Q_1^{eff}}{Q_t^{eff}} \gamma_t= \dfrac{1}{2} \dfrac{Q_1^{eff}}{Q_t^{eff}} \gamma_t
\label{eq:SE_S}.
\end{eqnarray}
Therein the actual Q-factor of the cavities is replaced by an effective quality factor $1/Q_{1,t}^{eff}=1/Q_{1,t}+1/Q_{em}$ taking into account the broad emitter linewidth $Q_{em}=\omega_0/\left(\gamma^*+\gamma_0\right)$. In the present case, we believe that even at the temperature of 77K which is used in the experiment, the linewidth of the emitter is still smaller or comparable to the cavity mode. Even in the case of strong dephasing the main conclusions of the paper are not affected, though the actual shape of the theoretical curves of the SE change in Fig.4 may slightly change. In any case the field delocalization and the change of the effective quality factor will determine the change of the decay rate of the emitter.

\subsection{\leftline{2. Sample growth and fabrication}}

The sample was grown by metal-organic vapor phase epitaxy on a InP (100) substrate misoriented $2^o$ toward the (110) facet. A 100 nm InP buffer layer was initially deposited, followed by a 110 nm InGaAsP layer with bandedge emission at 1250 nm. A single layer of QDs was then grown by depositing 2 monolayers (MLs) of InAs on top of a 1.2ML GaAs interlayer (Ref. \cite{Anantathanasarn2005} of the main text). A growth interruption in tertiary butyl arsine of 10 seconds was applied after the dot formation, which was followed by another 110 nm InGaAsP layer.  A 50 nm InP layer was finally grown as the capping layer. 

The photonic crystal cavities was fabricated using electron beam lithography and inductively coupled plasma etching with the Cl$_2$/Ar/H$_2$ mixture. The wet etching in HCl/H$_2$O was employed to remove the InP sacrificial layer and to make a 220 nm InGaAsP free standing membrane. A lattice constant of 480 nm and a filling factor of 0.30 were chosen to achieve the target cavity mode near 1550 nm. As seen from the $\mu$PL spectrum of the FP cavity in Fig. \ref{fig:Graph2}(c) of the main text, the target cavity mode is detuned ~15 nm from the dispersion edge of the waveguide mode supported by the FP cavity (see also Ref. \cite{Hoang2012} of the main text for a detailed discussion of the photonic band structure.)

\clearpage

\end{document}